\documentstyle[aps,prb]{revtex}

\begin{document}
\title{Frictional drag between non-equilibrium charged gases}
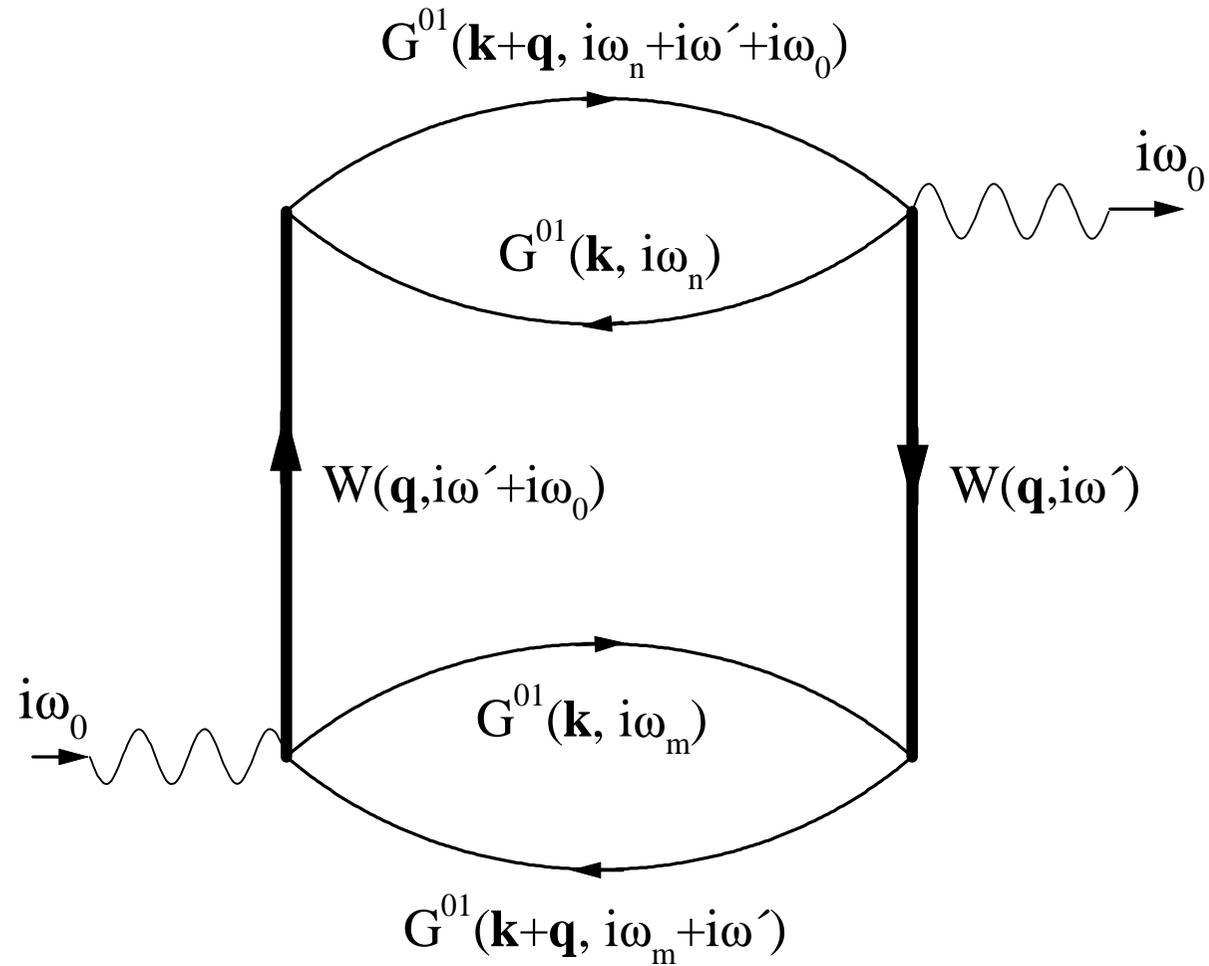
\author{X. F. Wang and I. C. da Cunha Lima}
\address{Instituto de F\'\i sica, Universidade do Estado do Rio de Janeiro, \\
Rua S\~{a}o Francisco Xavier 524, 20550-013 Rio de Janeiro, Brazil}
\date{\today}
\maketitle

\begin{abstract}
The frictional drag force between separated but coupled two-dimensional
electron gases of different temperatures is studied using the
non-equilibrium Green function method based on the separation of
center-of-mass and relative dynamics of electrons. As the mechanisms of
producing the frictional force we include the direct Coulomb interaction,
the interaction mediated via virtual and real TA and LA phonons, optic
phonons, plasmons, and TA and LA phonon-electron collective modes. We found
that, when the distance between the two electron gases is large, and at
intermediate temperature where plasmons and collective modes play the most
important role in the frictional drag, the possibility of having a
temperature difference between two subsystems modifies greatly the
transresistivity.
\end{abstract}

\section{Introduction}

\label{intro}

Decades after the original proposal,\cite{pogr} and after the first
successful measurement, \cite{solo} the frictional drag effect between
separated but coupled electron subsystems is still attracting a lot of
interest. In part this is due to the possibility of a direct view of
interactions between charged carriers, which play a key role in central
questions of condensed matter physics like superconductivity and quantum
Hall effects. After the initial theories based on the direct Coulomb drag
mechanism,\cite{gram1,jauho,zhen} electron-electron interactions mediated by
other channels, such as phonon exchange \cite{gram,flen,bons,tso,zhan,bada}
and electron-phonon collective modes,\cite{bons} have been explored as
sources of frictional drag at low temperatures. However, recent experiments
showed that further theoretical developments are still necessary before
obtaining a quantitative view of this problem\cite{noh}. At intermediate
temperatures, the importance of plasmons has been predicted theoretically by
Flensberg {\it et al}, \cite{flenplm} and confirmed experimentally by Hill 
{\it et al}, \cite{hill} but the discrepancies between theoretical and
experimental results are significant. G\"{u}ven and Tanatar \cite{guve} made
a detailed consideration of the optical phonon contribution to the plasmon
drag, and obtained an improvement on the theoretical result for the case
where optical phonons are of little importance on the drag force due to
difficulty of exciting them \cite{hu} at temperatures where plasmons
dominate. At the same time some studies concluded that the RPA should be
substituted by more detailed many-body theories \cite{stls1,stls2,stls3}
when estimating the frictional force. Besides those studies focusing on the
mechanism of frictional drag between 2D-2D electron-electron Fermi gas
subsystems, intensive theoretical \cite{stls2,stls3,dong,e-h,cui,tso1} and
experimental \cite{siva} works have been done on a 2D-2D electron-hole Fermi
gas, and also in other cases \cite{rojo,jorg,tana1,mag}.

Most works on Coulomb drag are made in the linear regime, in such a way that
the temperature is assumed to be the same for the two subsystems, and equal
to that of the heat bath. It is well-known, however, that hot electron
effects occur in semiconductor microstructures\cite{ivan}, where electrons
can be easily heated or cooled by very low electric voltage or photon
excitations. In non-equilibrium situations, as it is the case in a Coulomb
drag experiment, there is no reason to assume, from the very beginning, the
two system at the same temperature of the heat bath. The question of
thermalization of non-equilibrium systems is of interest also in several
other fields in physics \cite{zuba,kopo}. In this paper we study the
frictional drag in the non-equilibrium regime resulting from the direct
Coulomb interaction, the e-e interaction mediated by virtual and real TA and
LA phonons, optic phonons, plasmons, and TA and LA phonon-electron
collective modes. The frictional drag is described by a non-equilibrium
Green's function method, which has been applied successfully in the
development of quantum transport theory in semiconductors. We believe that
allowing in the calculation different temperatures for the two systems, what
is the central point in the present work, a new viewpoint is brought to this
developing area.

In the Lei-Ting formalism \cite{lei} used to describe non-linear transport
in semiconductor structures, the balance equations are established by
separating the center-of-mass from the relative motion. The Hamiltonian
becomes $H=H_{0}+H_{E}+H_{I}=H_{c}+H_{E}+H_{r}+H_{I}$, where $H_{c}$
corresponds to the motion of the center-of-mass, $H_{E}$ corresponds to the
time developing part due to external field, $H_{r}$ is the Hamiltonian of
relative electron system, and $H_{I}$ represents the scattering by
impurities and phonons. The last term is the one which causes the
dissipation of momentum and energy.

In order to obtain the equation of motion for the density matrix, we take as
initial state an equilibrium state of a system free of the interactions $%
H_{I}$ and $H_{E}$. The real system is obtained by adiabatically switching
on $H_{I}$ and $H_{E}$ at the infinite past time, till the present
non-equilibrium state. Then, the final state density matrix of the relative
electron system can be expanded in terms of the density matrix of the
initial state. If we choose a proper initial state close to the final state
of the relative electron system, a good result can be reached by keeping
only the term in first order of perturbation. This approach has made the
Lei-Ting balance equation an efficient tool to calculate the non-linear
transport properties in semiconductors, since the numerical calculation
involved is greatly simplified. However, this technique has not yet been
employed in the electron-electron frictional drag phenomena, mostly because
of the complexity of the many-body effect due to the inter-subsystem
electron-electron interaction. We noticed that the $H_{I}$ is important for
configuring the ensembles of the system for this problem, so higher order
terms must be considered. On the other hand, it is not convenient to include
those higher order terms in the usual version of the formalism. In this
paper, we propose a way to circumvent this difficulty in the case of the
frictional drag problem.

\section{Theory}

The Hamiltonian is written as$\ h=h_{0}+h_{I}$, with ensemble average
density matrices $\varrho ^{h}$ and $\varrho _{0}$ corresponding to $h$ and $%
h_{0}$, respectively. The ensemble average of an explicitly time dependent
Heisenberg operator $A_{t}$ is: 
\begin{equation}
\overline{A}_{t}=<\varrho ^{h}A_{t}>=Tr\{\varrho _{0}A_{t}\}-i\int_{-\infty
}^{t}dt_{1}Tr\{\varrho ^{h}[A_{t}^{h_{0}}(t-t_{1}),h_{It_{1}}^{h_{0}}]\},
\end{equation}
where, $O^{h}(t)=e^{iht}Oe^{-iht}$ and $%
O^{h_{0}}(t)=e^{ih_{0}t}Oe^{-ih_{0}t} $. Instead \ of approximating $\varrho
^{h}$ by $\varrho _{0}$, as in the standard version of Lei-Ting's balance
equations, we argue that, in cases where $h_{I}$ plays an important role in
the ensemble averaging process, it is preferable to use $A^{h}$ and $h^{h}$
in place of $A^{h_{0}}$ and $h^{h_{0}}$. The validity of this approximation
is the same as in the standard version, as we can see by considering the
Hamiltonian $h_{0}=h-h_{I} $ and expanding $\varrho _{0}$ in terms of $%
\varrho ^{h}$. The average value of a physical property $A_{t}$ becomes:

\begin{eqnarray}
\overline{A}_{t}=<\varrho ^{h}A_{t}> &=&Tr\{\varrho
_{0}A_{t}\}-i\int_{-\infty }^{t}dt_{1}Tr\{\varrho
^{h}[A_{t}^{h}(t-t_{1}),h_{It_{1}}^{h}]  \nonumber \\
&=&Tr\{\varrho _{0}A_{t}\}-i\int_{-\infty }^{t}dt_{1}Tr\{\varrho
^{h}[A_{t}^{h}(t),h_{It_{1}}^{h}(t_{1})]\}
\end{eqnarray}

Next, we consider a system of two coupled, but separated, quantum wells (no
particle exchange between them being allowed), with carrier densities \ and
electron temperatures $n_{1}$ and $T_{e1}$ in one well , and $n_{2}$ and $%
T_{e2}$ in the other. As a general case, it is assumed that the two
subsystems have drift velocities ${\bf v}_{1}$ and ${\bf v}_{2}$, and
applied electric fields ${\bf E}_{1}$ and ${\bf E}_{2}$, respectively. The
temperature of the phonon reservoir is $T_{p}$. After separating the
center-of-masses (located at positions ${\bf R}_{1}$ and ${\bf R}_{2}$) from
the relative coordinates, the Hamiltonian becomes: 
\begin{equation}
H=en_{1}{\bf R}_{1}{\bf \cdot E}_{1}+en_{2}{\bf R}_{2}{\bf \cdot E}%
_{2}+H_{1C}+H_{2C}+H_{e1}+H_{e2}+H_{e12}+H_{ph}+H_{e-ph}+H_{ei},
\end{equation}
where $H_{1C}=\frac{P_{1}^{2}}{2n_{1}m_{e}}$, $H_{2C}=\frac{P_{2}^{2}}{%
2n_{2}m_{e}}$, $H_{e1}=\sum_{{\bf k}}\varepsilon _{{\bf k}}c_{{\bf k}%
}^{\dagger }c_{{\bf k}}+\frac{1}{2}\sum_{{\bf q}_{1}}U_{C}(q_{1})\rho _{{\bf %
q}_{1}}^{1}\rho _{-{\bf q}_{1}}^{1}$, $H_{e2}=\sum_{{\bf p}}\varepsilon _{%
{\bf p}}d_{{\bf p}}^{\dagger }d_{{\bf p}}+\frac{1}{2}\sum_{{\bf q}%
_{2}}U_{C}(q_{2})\rho _{{\bf q}_{2}}^{2}\rho _{-{\bf q}_{2}}^{2}$, and $%
H_{e12}=\sum_{{\bf q}}U_{d}(q)\rho _{{\bf q}}^{1}\rho _{-{\bf q}}^{2}e^{i%
{\bf q}({\bf R}_{1}-{\bf R}_{2})}$. The density operator in system $1$ is $%
\rho _{{\bf q}}^{1}=\sum_{{\bf k}}c_{{\bf k+q}}^{\dagger }c_{{\bf k}}$, and,
for system $2$, $\rho _{{\bf q}}^{2}=\sum_{{\bf k}}d_{{\bf k+q}}^{\dagger
}d_{{\bf k}}$. The final three terms, $H_{ph}$, $H_{e-ph}$, and $H_{ei}$,
are defined as usual. \cite{lei} $U_{C}(q)$ and $U_{d}(q)$ are the intra-
and inter- well interactions. For the sake of simplifying the formalism, we
start by considering the interaction only through the Coulomb
electron-electron (e-e) potential.

As in the original Lei-Ting balance equation formalism, we choose an initial
equilibrium system composed of independent subsystems (electron gases in
well 1 and 2, and phonons) in the Hamiltonian $H_{r}=H_{e1}+H_{e2}+H_{ph}$,
with the initial density matrix $\varrho _{0}=\varrho _{e1}\varrho
_{e2}\varrho _{p}=e^{H_{p}/T_{p}}e^{H_{e2}/T_{e2}}e^{H_{e1}/T_{e1}}$. For
the density matrix of the final relative electrons of the Hamiltonian $%
H_{r}+H_{I}=H_{r}+H_{e12}+H_{e-ph}+H_{ei}$, we use $\tilde{\varrho}%
=e^{H_{p}/T_{p}}e^{H_{e2}/T_{e2}+H_{e1}/T_{e1}+H_{e12}/\sqrt{T_{e1}T_{e2}}}$%
, which we assume to be close to the steady state of the relative electrons.
The reason why we include $H_{e12}$ into the density matrix is that the e-e
interaction, which is very special and much stronger than other interactions
such as electron-impurity and electron-phonon interaction, should give
important contributions to the ensemble average for the final state of the
relative electrons. The present choice accelerates the convergency.

We consider the case where the electron temperature in well 2 is the same as
that of the lattice, $T_{p}=T_{e2}=T$, and $\gamma =T_{e1}/T$. By
introducing a new interaction representation corresponding to the virtual
Hamiltonian $\tilde{H}=\gamma H_{e1}+H_{e2}+H_{ph}+\sqrt{\gamma }H_{e12}$%
, where operators are expressed as $\tilde{O}(t)=e^{i\tilde{H}t}Oe^{-i%
\tilde{H}t}$, the Liouville equation to the first order becomes:

\begin{equation}
\varrho ^{h}=\varrho _{0}-i\int_{-\infty }^{t}dt_{1}e^{iH(t_{1}-t)}e^{-i%
\tilde{H}(t_{1}-t)}[\tilde{H}_{It_{1}}(t_{1}-t),\tilde{\varrho }%
]e^{i\tilde{H}(t_{1}-t)}e^{-iH(t_{1}-t)}
\end{equation}
By approximating $e^{iH(t_{1}-t)}e^{-i\tilde{H}(t_{1}-t)}\simeq
e^{-i(\gamma -1)H_{e1}(t_{1}-t)}$,\cite{lei} the ensemble average becomes:

\begin{equation}
\overline{A}_{t}=<\varrho ^{h}A_{t}>=Tr\{\varrho _{0}A_{t}\}-i\int_{-\infty
}^{\infty }dt_{1}\theta (t-t_{1})
Tr\{\tilde{\varrho }[\tilde{A}_{t}^{\gamma,(t_{1}-t)}(t),\tilde{H}_{It_{1}}(t_{1})]\}
\end{equation}
where $\tilde{A}_{t'}^{\gamma,\tau}(t)=
e^{i\tilde{H}t}e^{-i(\gamma -1)H_{e1}\tau}
A_{t'}e^{i(\gamma -1)H_{e1}\tau}e^{-i\tilde{H}t}$.

The time derivative operator of the total momentum for the electron gas in
well 2 is: 
\begin{equation}
{\bf \dot{P}}_{2}=-i[{\bf P}_{2},H]=-i\nabla _{{\bf R}_{2}}H=\sum_{{\bf q}}i%
{\bf q}U_{d}(q)\rho _{{\bf q}}^{1}\rho _{-{\bf q}}^{2}e^{-i{\bf q\cdot }(%
{\bf R}_{1}-{\bf R}_{2})}-en_{2}{\bf E}_{2}+{\bf \dot{P}}_{2,ep}+{\bf \dot{P}%
}_{2,ei}
\end{equation}
In the case where the circuit of the second well is opened (without
current), the phonon frictional force ${\bf \dot{%
P}}_{2,ep}$ and the impurity frictional force ${\bf \dot{P}}_{2,ei}$
disappear, together with the derivative of total moment of the second well.
In the subsequent equations we make ${\bf R}_{2}=0$.
The frictional force acting on the electron gas in the well 2 is balanced by
the electric force:

\begin{eqnarray}
en_{2}{\bf E}_{2} &=&i\sum_{{\bf q}}{\bf q}\int_{-\infty }^{\infty
}dt_{1}(-i)\theta (t-t_{1})\sum\limits_{{\bf k},{\bf p}}\sum\limits_{{\bf Q}%
}e^{-i{\bf q\cdot }[{\bf R}_{1}(t)-{\bf R}_{1}(t_{1})]+i{\bf Q\cdot R}%
_{1}(t_{1})}e^{i(1-\gamma )(\varepsilon _{k+q}-\varepsilon _{k})(t-t_{1})} 
\nonumber \\
&&\times Tr\left\{ \tilde{\varrho }[U_{d}(q)\tilde{c}_{{\bf k}+{\bf q}%
}^{\dagger }(t)\tilde{c}_{{\bf k}}(t)\tilde{d}_{{\bf p}-{\bf q}%
}^{\dagger }(t)\tilde{d}_{{\bf p}}(t),U_{d}({\bf Q}-{\bf q})\tilde{%
\rho }_{{\bf Q}-{\bf q}}^{1}(t_{1})\tilde{\rho }_{{\bf q}-{\bf Q}%
}^{2}(t_{1})]\right\} .
\end{eqnarray}

The above expression can be simplified if (i) the velocity fluctuations are
neglected by making ${\bf R}_{1}(\tau )={\bf v}_{d}\tau $, where ${\bf v}%
_{d} $ is the drift velocity of the electron gas in well 1, and (ii) we
consider zero frequency drag, where the imaginary part of the frictional
force is zero. Furthermore, the system satisfies the time invariance
transformation in steady state with ${\bf Q}=0$. Then we have

\begin{equation}
en_{2}{\bf E}_{2}=-Im \sum_{{\bf k},{\bf p},{\bf q}}{\bf q}\Gamma ^{R}({\bf k%
},{\bf p},{\bf q},\omega _{0})
\end{equation}
where $\omega _{0}=qv_{d}+(1-\gamma )(\varepsilon _{{\bf k}+{\bf q}%
}-\varepsilon _{{\bf k}})$, and $\Gamma ^{R}({\bf k},{\bf p},{\bf q},\omega
) $ is the Fourier transform of the retarded Green's function 
\begin{equation}
\Gamma ^{R}({\bf k},{\bf p},{\bf q},\tau )=-i\theta (\tau )Tr\left\{ 
\tilde{\varrho }[U_{d}(q)\tilde{c}_{{\bf k}+{\bf q}}^{\dagger }(\tau )%
\tilde{c}_{{\bf k}}(\tau )\tilde{d}_{{\bf p}-{\bf q}}^{\dagger }(\tau )%
\tilde{d}_{{\bf p}}(\tau ),U_{{\bf d}}(-q)\tilde{\rho }_{-{\bf q}%
}^{1}(0)\tilde{\rho }_{{\bf q}}^{2}(0)]\right\}
\end{equation}
The later can be obtained by analytical continuation from its Matsubara
counterpart:

\begin{equation}
\Gamma ^{R}({\bf k},{\bf p},{\bf q},\omega )=\lim_{i\omega _{n}\rightarrow
\omega +i\delta ^{+}}\Gamma ^{M}({\bf k},{\bf p},{\bf q},i\omega _{n})
\end{equation}

After a series of calculation to expand $\Gamma ^{M}({\bf k}, {\bf p},{\bf q}%
,i\omega _{n})$ in the equilibrium representation without inter-subsystem
interaction $H_{I}$, we obtain a diagram, where the renormalization is made
with the polarization function of well 1 and well 2, rather than
inter-subsystem interaction. This diagram is similar to the result of the
Memory Function method, but with different electron temperature in different
subsystems. Following the suggestion of L. Zheng and A. H. MacDonald, \cite
{zhen} we adopt the renormalization of the Coulombic interaction between
electron gases 1 and 2, as shown in Fig. 1. The drag resistivity then
becomes:

\begin{eqnarray}
R_{d}=\frac{E_{2}}{en_{1}v_{d}}=-\frac{1}{\pi n_{1}n_{2}e^{2}v_{d}}%
\sum\limits_{{\bf q}}q_{x}\int_{-\infty }^{\infty }d\omega &&\left| \frac{%
U_{d}(q)}{\epsilon _{T}({\bf q},\omega )}\right| ^{2}\hat{\Pi}_{2}^{1}({\bf q%
},\omega ,T_{1})\hat{\Pi}_{2}^{2}(q,\omega +\omega _{0},T)  \nonumber \\
&&\left[ n\left( \frac{\omega +\omega _{0}}{T}\right) -n\left( \frac{\omega 
}{T_{1}}\right) \right] .
\end{eqnarray}
The drift velocity of electrons in well 1 is assumed along the $x$
direction. ${\bf q}$ is the carrier momentum, $\omega $ is the carrier
energy, $n(x)$ is the Boson-Einstein distribution function, ${\bf \hat{\Pi}}%
^{1}({\bf q},\omega ,T_{1})=\hat{\Pi}_{1}^{1}({\bf q},\omega ,T_{1})+i\hat{%
\Pi}_{2}^{1}({\bf q},\omega ,T_{1})$, and ${\bf \hat{\Pi}}^{2}({\bf q}%
,\omega ,T)=\hat{\Pi}_{1}^{2}({\bf q},\omega ,T)+i\hat{\Pi}_{2}^{2}({\bf q}%
,\omega ,T)$, are the zero-order polarization functions of carriers in well 1 and well 2,
respectively. The dielectric function of the electron system is 
\begin{equation}
\epsilon _{T}({\bf q},\omega )=[1-U_{C}(q){\bf \hat{\Pi}}^{1}({\bf q},\omega
,T_{1})][1-U_{C}(q){\bf \hat{\Pi}}^{2}({\bf q},\omega ,T_{2})]-U_{d}^{2}(q)%
{\bf \hat{\Pi}}^{1}({\bf q},\omega ,T_{1}){\bf \hat{\Pi}}^{2}({\bf q},\omega
,T_{2})
\end{equation}

These expressions were obtained by assuming that only the Coulomb
interaction is important to the electronic coupling. It is easy to extend
this result to the case in which other mechanisms of interaction between
electrons become important, just by replacing the inter- and intra-well e-e
Coulomb potential by the sum of the corresponding interaction potentials due
to different mechanisms. For instance, if phonon mediated electron-electron
interaction is to be taken into account, the following substitution should
be made:

\begin{eqnarray}
U_{C}(q) &\rightarrow &U_{ii}({\bf q},\omega )=U_{C}(q)+\sum_{\lambda
}D_{ii}^{\lambda }({\bf q},\omega )  \nonumber \\
U_{d}(q) &\rightarrow &U_{12}({\bf q},\omega )=U_{d}(q)+\sum_{\lambda
}D_{12}^{\lambda }({\bf q},\omega )
\end{eqnarray}

Here $D_{ij}^{\lambda }=\int \frac{dq_{z}}{2\pi }|M^{\lambda }({\bf q}%
,\omega )|^{2}f_{i}(q_{z})f_{j}(q_{z})[\frac{1}{\omega -\omega _{\lambda
}+i/2\tau _{ph}^{\lambda }}-\frac{1}{\omega -\omega _{\lambda }-i/2\tau
_{ph}^{\lambda }}]$ is the interaction potential between electrons in wells $%
i$ and $j$ mediated by phonons of the branch $\lambda $, as discussed in Ref.%
\onlinecite{flen}. $M^{\lambda }({\bf q},\omega )$ is the bulk matrix of
electron-phonon interaction, and $f_{i}(q_{z})$ is the form factor for the
electron-phonon interaction in well $i$. The mean free time of phonons $\tau
_{ph}^{\lambda }$ equals the mean free path $l_{ph}^{\lambda }$ divided by
phonon velocity $v_{\lambda }$. For small drift velocities, the drag
resistivity becomes:

\begin{eqnarray}
R_{d} &=&\frac{1}{2\pi ^{3}e^{2}n_{1}n_{2}}\int_{0}^{\pi /2}d\theta \cos
^{2}\theta \int_{0}^{\infty }q^{3}dq\int_{0}^{\infty }d\omega \left| \frac{%
U_{12}({\bf q},\omega )}{\epsilon _{T}({\bf q},\omega )}\right| ^{2}\hat{\Pi}%
_{2}^{1}({\bf q},\omega ,T)  \nonumber \\
&&\times \left\{ \frac{\hat{\Pi}_{2}^{2}({\bf q},\omega ,T_{2})}{T_{2}\sinh
^{2}(\omega /2T_{2})}+2\left[ \coth (\omega /2T_{1})-\coth (\omega /2T_{2})%
\right] \frac{\partial }{\partial \omega }\hat{\Pi}_{2}^{2}({\bf q},\omega
,T_{2})\right\} .  \label{tr}
\end{eqnarray}
The real (virtual) phonon contribution comes from the imaginary (real) part
of $D_{ij}^{\lambda }$ and the contribution of plasmons and electron-phonon
collective modes come from the zeroes of the real part of $\epsilon _{T}(%
{\bf q},\omega )$. The well-know expression for the linear drag resistivity
is recovered if we neglect the anisotropy of the interaction matrix, and
assume the same temperature $T_{1}=T_{2}=T$\ for the electron gases in both
wells. Then, the drag resistivity of the two-temperature model becomes the
generally accepted linear result:

\begin{equation}
R_{d}{\normalsize =\frac{1}{8\pi ^{2}e^{2}n_{1}n_{2}T}}\int_{0}^{\infty }dq%
{\normalsize q^{3}\int_{0}^{\infty }d\omega \left| \frac{U_{12}(q,\omega )}{%
\epsilon _{T}(q,\omega )}\right| ^{2}}\frac{{\normalsize \hat{\Pi}%
_{2}^{1}(q,\omega )\hat{\Pi}_{2}^{2}(q,\omega )}}{\sinh ^{2}(\omega /2T)}%
{\normalsize ,}
\end{equation}

\section{Results and Conclusion}

The numerical calculation is performed in a model in which the electron
system occupies two GaAs quantum wells separated by infinite barriers. This
model may underestimate the electron effective mass and the correlation
between two wells, but these effects are negligible for the cases discussed
in this paper. In addition, it is well known that the RPA is expected to
give poor results for low density electron gases, and several improvements
beyond RPA have been employed in the literature, as already discussed in
Chap. \ref{intro}. However, RPA is used in this paper for estimating the
many-body effects, since the mechanism leading to the unexpected large
measured drag resistivity has not yet been clarified, and also because our
aim here is not focused on the improvements or more detailed estimative or
corrections beyond RPA. We believe that local field corrections to our
result give similar results as discussed in other places. It is worthwhile
to mention that effects of the optic phonon directly causing, or mediating,
the frictional drag were considered in Refs. \onlinecite{guve} and %
\onlinecite{flen}. Here the optic phonon contribution is taken into account
just by using the static dielectric constant $\kappa _{s}$ as the background
dielectric function.\cite{card} An argument in favor of this assumption is
that the plasmon energy is much smaller than those of optic phonons, and at
the plasma frequency the optic phonon contribution to dielectric function is
practically frequency independent.

The contributions of the longitudinal acoustic (LA) and the transversal
acoustic (TA) phonons via deformation potential and piezoelectric
interaction, on the other hand, are fully included. The mean free times, $%
\tau _{^{LO}ph}$ and $\tau _{^{TO}ph}$, are used as fitting parameters for
both the LA and the TA branches. The parameters used for GaAs are $\kappa
_{s}=12.9$, $d_{1}=5.31g/cm^{3}$, $\Xi =14eV$, $m^{\ast }=0.067$, $%
v_{sl}=5.29\times 10^{3}m/s$, $e_{14}=1.41\times 10^{19}V/m$, and $%
v_{st}=2.48\times 10^{3}$ for the dielectric constant, density, deformation
potential, electron effective mass, longitudinal sound velocity,
piezoelectric constant and transversal sound velocity, respectively. The
numerical calculation is performed for a structure of two $200$\AA\ wide
quantum wells, with a $300$\AA\ barrier between them.

In Fig.2 (a), the drag resistivity $R_{d}$ \ is plotted as a function of the
equilibrium temperature in ($T_{1}=T_{2})$ for a matching electron densities
of $n=1.37\times 10^{11}cm^{-2}$ in both wells. The calculation is performed
in a range of values of $\tau _{_{^{LA}}}$ and $\tau _{_{^{TA}}}$\ suitable
for analyzing the relative importance of the electron-phonon collective mode
for the LA and TA phonons. The solid lines represent results of different
values of the phonon mean free paths (the same for LA and TA phonons),
namely $l_{ph}=10^{-2},1,10^{2}cm$. \ In the two upper curves ($1cm$,$%
10^{2}cm$) the electron-phonon collective modes dominate the contribution.
In the lower curve ($10^{-2}cm$), however the main contribution comes from
the virtual and real phonon terms. Similarly to previous studies\cite
{bons,hill}, our theoretical results fit qualitatively well the experimental
data from Ref. \onlinecite{hill} (filled circles), but several quantitative
inconsistencies still remain. For instance, at low temperatures the
calculated peak of the acoustic phonon is displaced to the high temperature
side. If we allow $l_{LA}$ and $l_{TA}$ to be different and diminish the
value of $l_{LA}$ to reduce the contribution of LA phonons, the peak moves
to the low temperature side. In dotted line, we show the drag resistivity
including only the contributions of \ the TA phonons ($l_{TA}=10^{2}cm$),
together with the Coulomb interaction. Comparing with the experimental
result, this seems to indicate that TA phonons have a longer mean free path
then LA phonon, which can result in the dominance of TA phonons in the
frictional drag force as mentioned in Ref. \cite{bada}. However, even though
the position of the calculated peak can coincide with the experimentally
observed by fitting $l_{LA}$ and $l_{TA}$, the value of the maximum cannot
reach the experimental result in this low carrier density case. At
intermediate temperatures ($T\gtrsim 0.2T_{F}$), the plasmon begin to
dominate the contribution to frictional drag force, which reaches its
maximum at about ($T\simeq 0.5T_{F})$, as calculated in Ref. \cite{hill}.
Our conclusion is that the LA phonons enhance the plasmon peak at
intermediate temperatures, in the same way as the optic phonon, \cite{guve},
while the TA phonons have little influence at this temperature range.
However, our calculation gives a smaller frictional drag force, and a higher
peak temperature than what is observed experimentally. In Fig. 2(b), we show
the dependence of the drag resistivity on the temperature, as in Fig. 2(a),
with a different carrier density, $n=2.66\times 10^{11}cm^{-2}$, and several
TA and LA mean free paths, namely, $l_{ph}=10^{-5},10^{-2},1cm$. Differently
from the previous low density regime, we find a very short phonon mean free
path, $l_{ph}=0.01cm$, when fitting the experimental result. Since samples
differing just by the carrier densities are expected to have the same order
of the phonon mean free path, this points to a disagreement between
theoretical and experimental results for the density dependence of the
frictional drag force. If the LA phonon is excluded from the calculation, as
denoted by the dotted curve, even a very long mean free path for TA phonon, $%
l_{TA}=100cm$, results in a very small frictional drag force.

Next, we relax the condition $T_{1}=T_{2}$, allowing for the driving well to
have a different temperature from that of the drag well. In Fig. 3(a) and
(b) we calculate the dependence of the drag resistivity on the normalized
temperature divided by the Fermi temperature of the drag well, $T_{2}/T_{F}$%
. Electrons in the drag well are assumed in thermal equilibrium with the
phonons. Curves are plotted for different ratios $T_{1}/T_{2}$\ . It is
found that a small change of that temperature ratio affects little the
phonon drag peak at low temperatures, but modifies greatly the plasmon peak
at intermediate temperatures. When the driving electrons are cooler than the
drag electrons, the frictional drag force increases greatly. In the reverse
case the plasmon peak diminishes, or even disappears. If we take a detailed
look at the frictional drag, given by Eq.(\ref{tr}{\it )}, we find that the
changes introduced by non-equilibrium come mainly from the second term. It
reflects the modification of the correlation function by the drift energy of
electrons in the driving well, and it is of second order in the equilibrium.

In Fig. 4(a) we calculate the non-equilibrium drag resistivity divided by
that obtained in equilibrium, as a function of the temperature ratio $%
T_{1}/T_{2}$, at various $T_{2}$, in the case of matching densities $%
n_{1}=n_{2}=2.66\times 10^{11}cm^{-2}$. At low temperatures ($3K$), a small
change on the temperature of the driving electron gas makes little effect on
the value $R_{d}T_{2}/R_{d}^{0}T_{1}$, as shown by the almost flat solid
curve at $R_{d}T_{2}/R_{d}^{0}T_{1}=1$. At higher temperatures, however,
when the interaction with plasmons becomes important, a cooler (hotter)
driving electron gas enhances (reduces) greatly the drag resistivity. The
derivatives of curves in Fig.4(a) at the point $T_{1}=T_{2}$ are plotted \
in Fig. 4(b) as functions of the temperature of the drag well. Results are
also shown for the matching electron density $n=1.37\times 10^{11}cm^{-2}$.
The sensitivity of the plasmon contribution to the temperature difference $%
\delta T=T_{1}-T_{2}$ is a direct result of the non-linearity of the
correlation function over frequency, confirming that a correct approach to
the Coulomb drag effect must contain the non-equilibrium assertive of the
present work.

In order to make the above point clearer, we obtain, from Eq. (\ref{tr}),
the expression for the resistivity difference $\delta R_{d}$ resulting from
a temperature difference near $T_{2}=T$.

\begin{equation}
\delta R_{d}=\frac{\delta T}{T}{\normalsize \frac{1}{8\pi
^{2}e^{2}n_{1}n_{2}T}}\int_{0}^{\infty }{\normalsize dqq^{3}\int_{0}^{\infty
}d\omega \left| W_{12}(q,\omega )\right| ^{2}}\frac{{\normalsize \hat{\Pi}%
_{2}^{1}(q,\omega )\hat{\Pi}_{2}^{2}(q,\omega )}}{\sinh ^{2}(\omega /2T)}%
\Theta (q,\omega ),
\end{equation}
where $\Theta (q,\omega )=\frac{\omega }{\hat{\Pi}_{2}^{2}(q,\omega )}%
\left. \frac{\partial }{\partial \varpi }\hat{\Pi}_{2}^{2}(q,\varpi
)\right| _{\varpi =\omega }$. If $\hat{\Pi}_{2}^{2}(q,\omega )$ is a
linear function of $\omega $, then $\Theta (q,\omega )=1$ and $\delta R_{d}=%
\frac{\delta T}{T}R_{d}$, i.e. $R_{d}/(T_{1}T_{2})$ is a constant near $%
T_{1}=T_{2}$. Otherwise, $R_{d}/(T_{1}T_{2})$ will be sensitive to the
temperature difference. The contribution from acoustic phonons comes mainly
from $q\simeq 2k_{F}$, where $\hat{\Pi}_{2}^{1}(q,\omega )$ is a linear
function of $\omega $ in the range of the corresponding phonon energy (with
the parameters used in this paper). This leads to the fact that the acoustic
peak in Fig. 3 does not show any non-equilibrium trace. On the other hand,
the plasmon energy is higher than the energy of the single particle
excitation, and the corresponding non-linearity of $\hat{\Pi}%
_{2}^{1}(q,\omega )$ invokes strong non-equilibrium character for drag
resistivity $R_{d}/(T_{1}T_{2})$ due to plasmons.

Fig. 5 shows the drag resistivity $R_{d}/(T_{1}T_{2})$ in equilibrium cases
(solid curves) as a function of the ratio $n_{1}/n_{2}$ of the electron
densities at different drag temperatures, $T_{2}=0.04$, $0.09$, $0.16$, $0.24
$, and $0.44T_{F}$ in comparison with the experimental results. The mean
free paths of the LA and TA phonons is assumed to be $l_{ph}=0.01cm$. We
found that the acoustic phonon enhances the drag resistivity at large $%
n_{1}/n_{2}$. It can be expected that the inclusion of local field effects
will provide good fittings between theoretical and experimental results,\cite
{hill} except at temperature $T_{2}=0.44T_{F}$. The results with bare
Coulomb interaction is shown in equilibrium cases by dotted curves, and in
non-equilibrium cases with $T_{1}=0.9T_{2}$ by dashed curves.

In conclusion, we have studied the momentum transfer rate between two nearby
separated electron gases, which are coupled via the Coulomb interaction and
phonon mediated interaction. We focused on the non-equilibrium
configuration, and a theory was developed to describe the frictional drag
force felt by one electron gas as a result of the relative drift of the
other, taking into account the possibility of difference on the subsystems
temperatures. It was found that a cooler (hotter) driving electron gas
greatly enhances (decreases) the frictional force caused by plasmons. This
behavior results from the fact the plasmons locate in a region of the  $%
\omega -k$ space where the  correlation function shows a nonlinear frequency
dependence, and they modifies little the acoustic phonons force because the
latter are important  mainly in the region of the  energy-momentum space
where the correlation function is  linear with the  frequency. 

\bigskip 

\acknowledgements
We thank X.L. Lei and B. Dong for fruitful discussions.

\begin{figure}[tbh]
\caption{Feymann diagram for the drag resistivity}
\label{fig01}
\end{figure}

\begin{figure}[tbh]
\caption{Drag resistivity $R_{d}/T^{2}$ as functions of temperature $T/T_{F}$
at electron density (a) $n=1.37\times 10^{11}cm^{-2}$ ($T_F=57K$), and (b) $%
n=2.66\times 10^{11}cm^{-2}$ ($T_F=110K$). Results including the direct
Coulomb interaction plus the LA and TA phonons with the same mean free paths 
$l_{ph}$ are plotted by solid lines. Experimental results according to Ref. 
{\protect{\onlinecite{hill}}} are shown as filled circles. The dashed line shows the
results when LA phonons are excluded in the calculation to better identify
the peak position of TA phonon. The phonon mean free paths are denoted by
numbers besides the curves in $cm$.}
\label{fig02(a)}
\end{figure}

\begin{figure}[tbh]
\caption{ Drag resistivity as functions of drag temperature $T_{2}$ at
different driving/drag temperature ratio $T_1/T_2$, denoted by numbers
besides the curves. The electron densities and phonon mean free paths are
(a) $n=1.37\times 10^{11}cm^{-2}$, $l_{ph}=100 cm$ and (b) $n=2.66\times
10^{11}cm^{-2}$, $l_{ph}=0.01 cm$. Experimental results from Ref. 
{\protect{\onlinecite{hill}}} are swhon as filled circles.}
\label{fig03}
\end{figure}

\begin{figure}[tbh]
\caption{ (a) The normalized drag resistivity $R_{d}T_{2}/R_{d}^{0}T_{1}$ as
functions of driving/drag temperature ratio $T_{1}/T_{2}$ at fixed drag
temperature $T_2$ numbered besides curves, and (b) its derivative at $%
T_{1}/T_{2}=1$. Only the direct Coulomb interaction is considered.}
\label{fig04}
\end{figure}

\begin{figure}[tbh]
\caption{ Drag resistivity $R_{d}/(T_{1}T_{2})$ is shown as function of
driving/drag electron density $n_{1}/n_{2}$ at different normalized drag
temperature $T_{2}/T_{F}=0.04,0.09,0.16,0.24,0.44$ and driving electron
density $n_{1}=2.66\times 10^{11}cm^{-2}$. Experimental results from Ref. 
\protect\cite{hill} is shown by filled circles. Solid lines represent
results including contributions from all kinds of mechanisms in equilibrium,
with TA and LA phonon mean free paths $l_{ph}=0.01cm$. Dotted lines
represents direct Coulomb drag forces (including plasmons) in equilibrium.
Dashed lines show the Coulomb drag forces in non-equilibrium cases with $%
T_1/T_2=0.9$}
\label{fig05}
\end{figure}


\begin{references}
\bibitem{pogr}  M. B. Pogrebinskii, Fiz. Tekh. Poluprovodn. {\bf 11}, 637
(1977) [Sov. Phys. Semicond. 11, 372 (1977)]. P. J. Price, Physica B\&C {\bf %
117}, 750 (1983).

\bibitem{solo}  P. M. Solomon, P. J. Price, D. J. Frank and D. C. La Tulipe,
Phys. Rev. Lett. 63, 2508 (1989).

\bibitem{gram1}  T. J. Gramila, et al., Phys. Rev. Lett.{\bf \ 66},
1216(1991).

\bibitem{gram}  T. J. Gramila, et al., Phys. Rev. B {\bf 47}, 12957(1993).

\bibitem{jauho}  A. P. Jauho and H. Smith, Phys. Rev. B{\bf 47}, 4420(1993).

\bibitem{zhen}  L. Zheng and A. H. MacDonald, Phys. Rev. B {\bf 48}, 8203
(1993).

\bibitem{flen}  K. Flensberg, et al., Phys. Rev. B {\bf 52}, 14761(1995).

\bibitem{bons}  M. C. Bonsager, K. Flensberg, B. Y. K. Hu and A. H.
MacDonald, Phys. Rev. B{\bf \ 57}, 7085 (1998);

\bibitem{tso}  H. C. Tso, P. Vasilopoulos and F. M. Peeters, Phys. Rev.
Lett. {\bf 68}, 2516(1992). H. C. Tso and P. Vasilopoulos, Phys. Rev. B {\bf %
45}, 1333(1992). H. C. Tso and N. J. M. Horing, Phys. Rev. B {\bf 44},
8886(1991).

\bibitem{zhan}  C. Zhang and Y. Takahashi, J. Phys.: CM {\bf 5}, 5009(1993).

\bibitem{bada}  S. M. Badalyan and U. R\"{o}ssler, Phys. Rev. B{\bf 59},
5643(1999).

\bibitem{noh}  H. Noh, et al., Phys. Rev. B {\bf 59}, 13114 (1999); ibid. 
{\bf 58}, 12621(1998).

\bibitem{flenplm}  K. Flensberg and B. Y. K. Hu, Phys. Rev. Lett. {\bf 73},
3572(1994); Phys. Rev. B {\bf 52}, 14796(1995).

\bibitem{hill}  N. P. R. Hill, et al., Phys. Rev. B {\bf 78}, 2204 (1997).

\bibitem{guve}  K. G\"{u}ven and B. Tanatar, Phys. Rev. B{\bf \ 56}, 7535
(1997).

\bibitem{hu}  B. Y. K. Hu, Phys. Rev. B{\bf \ 57}, 12345(1998).

\bibitem{stls1}  B. Tanatar and C. Bulutay, Phys. Rev. B {\bf 59},
15019(1999)

\bibitem{stls2}  I. V. Gornyi, A. G. Yashenkin and D. V. Khveshchenko, Phys.
Rev. Lett. {\bf 83}, 152(1999).

\bibitem{stls3}  L. Zheng and A. H. MacDonald, Phys. Rev. B {\bf 49},
5522(1994).

\bibitem{dong}  B. Dong and X. L. Lei, Eur. Phys. J. B {\bf 7}, 147(1999).

\bibitem{e-h}  L. \'{S}wierkowski, J. Szyma'{n}ski, and Z. W. Gortel, Phys.
Rev. Lett. {\bf 74}, 3245(1995); J. Szyma\'{n}ski, L. \'{S}wierkowski and D.
Neilson, Phys. Rev. B {\bf 50}, 11002(1994).

\bibitem{cui}  H. L. Cui, X. L. Lei and N. J. M. Horing, Superlattices and
Microstructures, {\bf 13}, 221(1993).

\bibitem{tso1}  H. C. Tso, P. Vasilopoulos and F. M. Peeters, Phys. Rev.
Lett. {\bf 70}, 2146 (1993).

\bibitem{siva}  U. Sivan, P. M. Solomon and H. Shtrikman, Phys. Rev. Lett. 
{\bf 68}, 1196(1992).

\bibitem{rojo}  A. G. Rojo, J. Phys.: CM {\bf 11}, R31 (1999).

\bibitem{jorg}  C. J\"{o}rger, et al., Phys. Rev. B {\bf 62}, 1572 (2000).

\bibitem{tana1}  B. Tanatar, Phys. Rev. B58, 1154(1998); B. Tanatar and A.
K. Das, ibid. {\bf 61}, 15959(2000);

\bibitem{mag}  M. W. Wu, H. L. Cui and N. J. M. Horing, Mod. Phys. Lett. B%
{\bf 10}, 279(1996); H. C. Tso and Vasilopoulos, Phys. Rev. B {\bf 4}5,
1333(1992).

\bibitem{ivan}  I. C. da Cunha Lima, X. F. Wang and X. L. Lei,\ Phys. Rev. B 
{\bf 55}, 10681(1997); X. F. Wang and X. L. Lei,{\it \ ibid.} {\bf 49},
4780(1994).

\bibitem{zuba}  D. N. Zubarev, {\it Non-equilibrium Statistical
Thermodynamics}, Consultants Bureau, New York, 1974.

\bibitem{kopo}  I. Koponen, Phys. Rev. E {\bf 55}, 7759 (1997).

\bibitem{lei}  X.L. Lei, and C.S. Ting, Phys. Rev. B {\bf 32}, 1112 (1985);
X.L. Lei and N.J.M. Horing, Int. J. Mod. Phys. B {\bf 6}, 805 (1992).

\bibitem{card}  Peter Y. Yu and Manuel Cardona, {\it Fundamentals of
Semiconductors}, Springer-Verlag, Berlin, 1996.
\end{references}
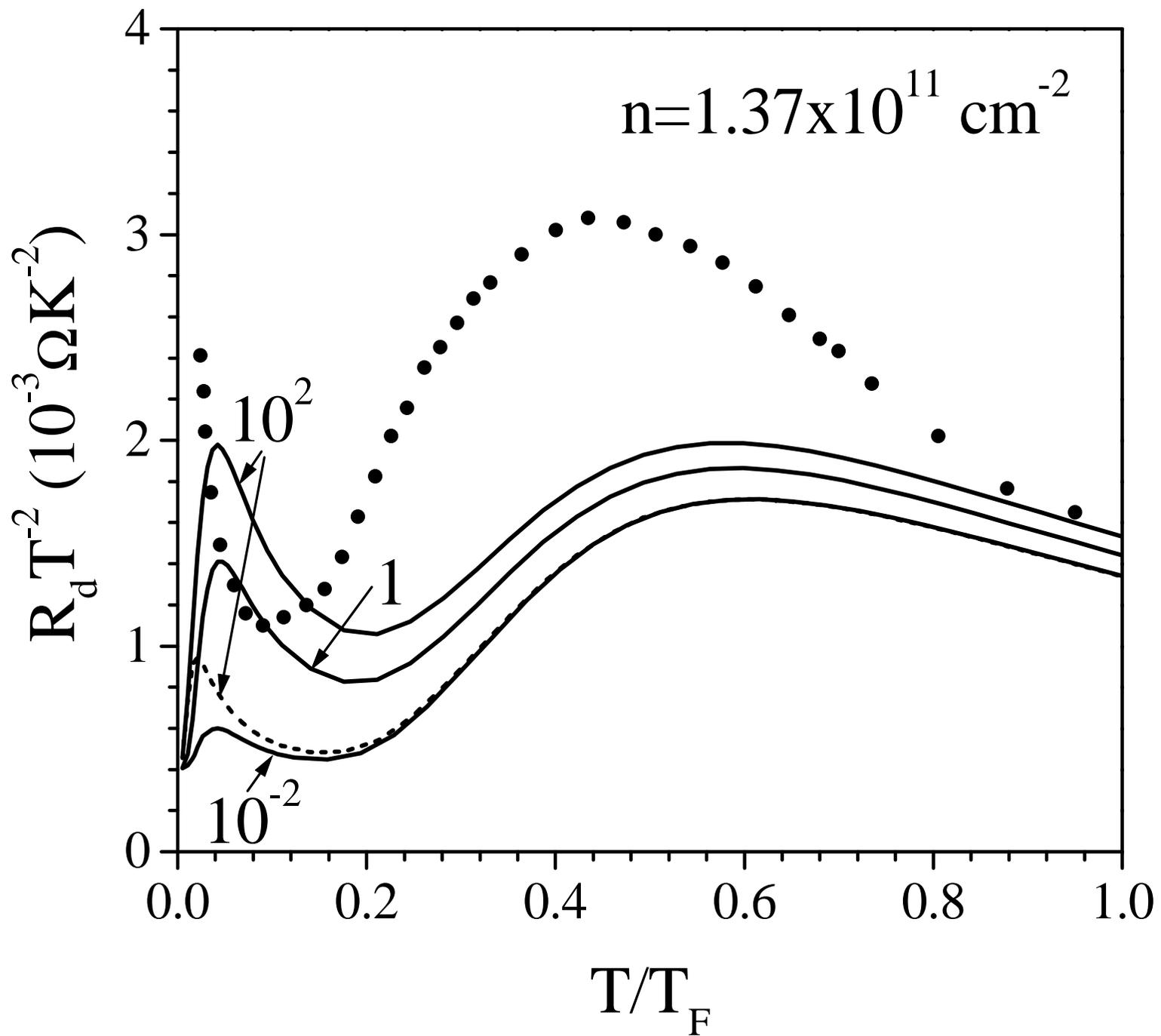
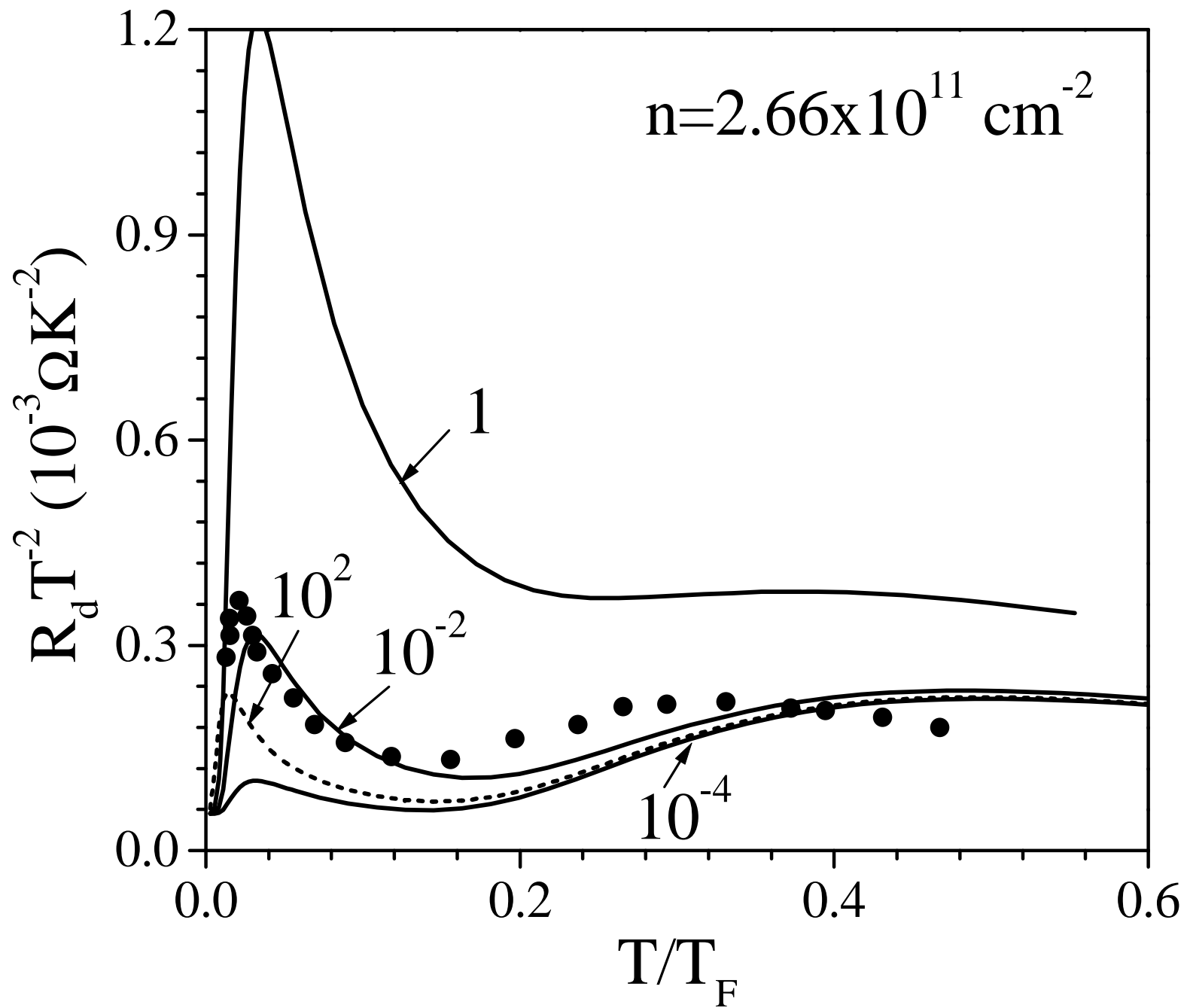
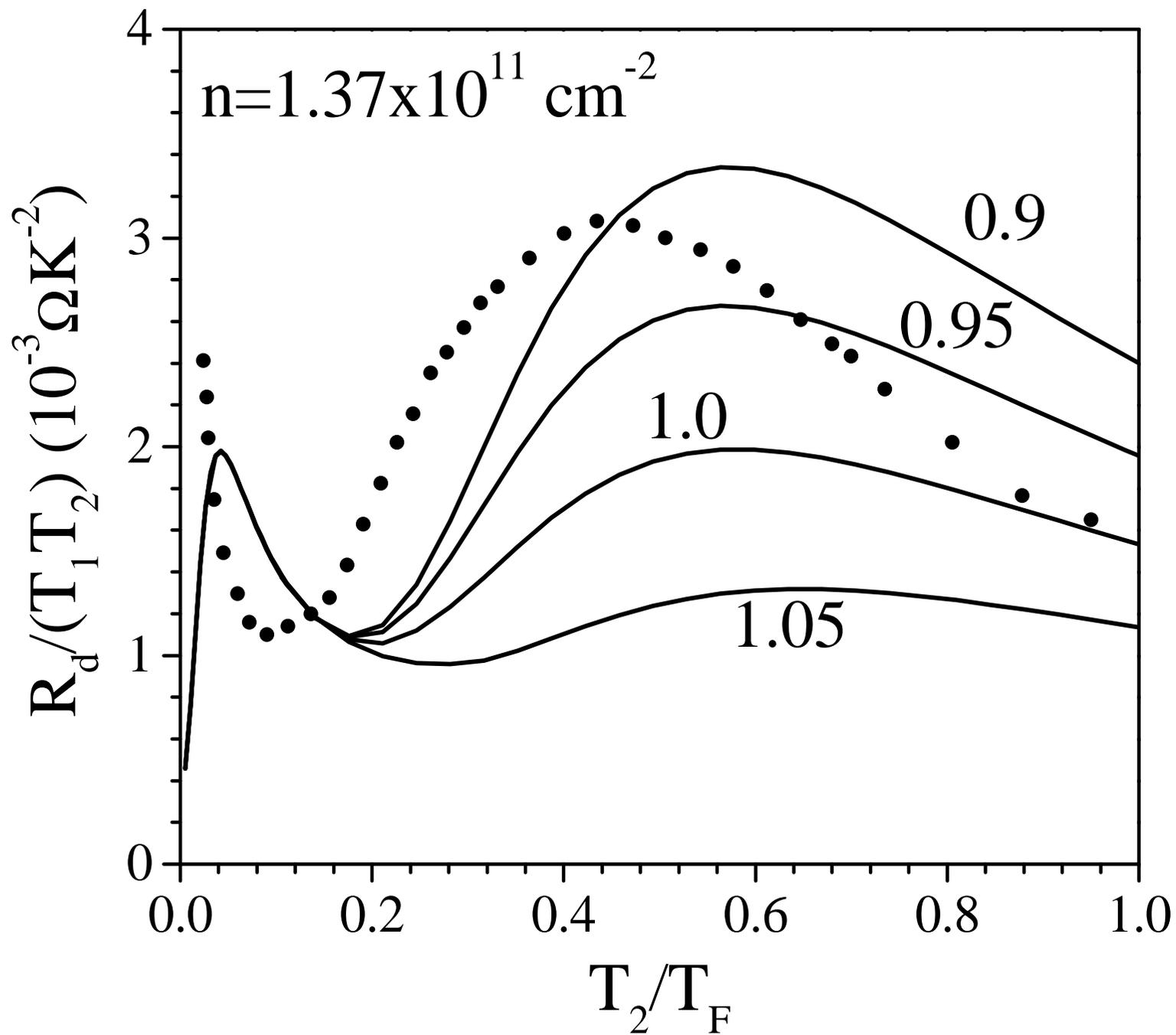
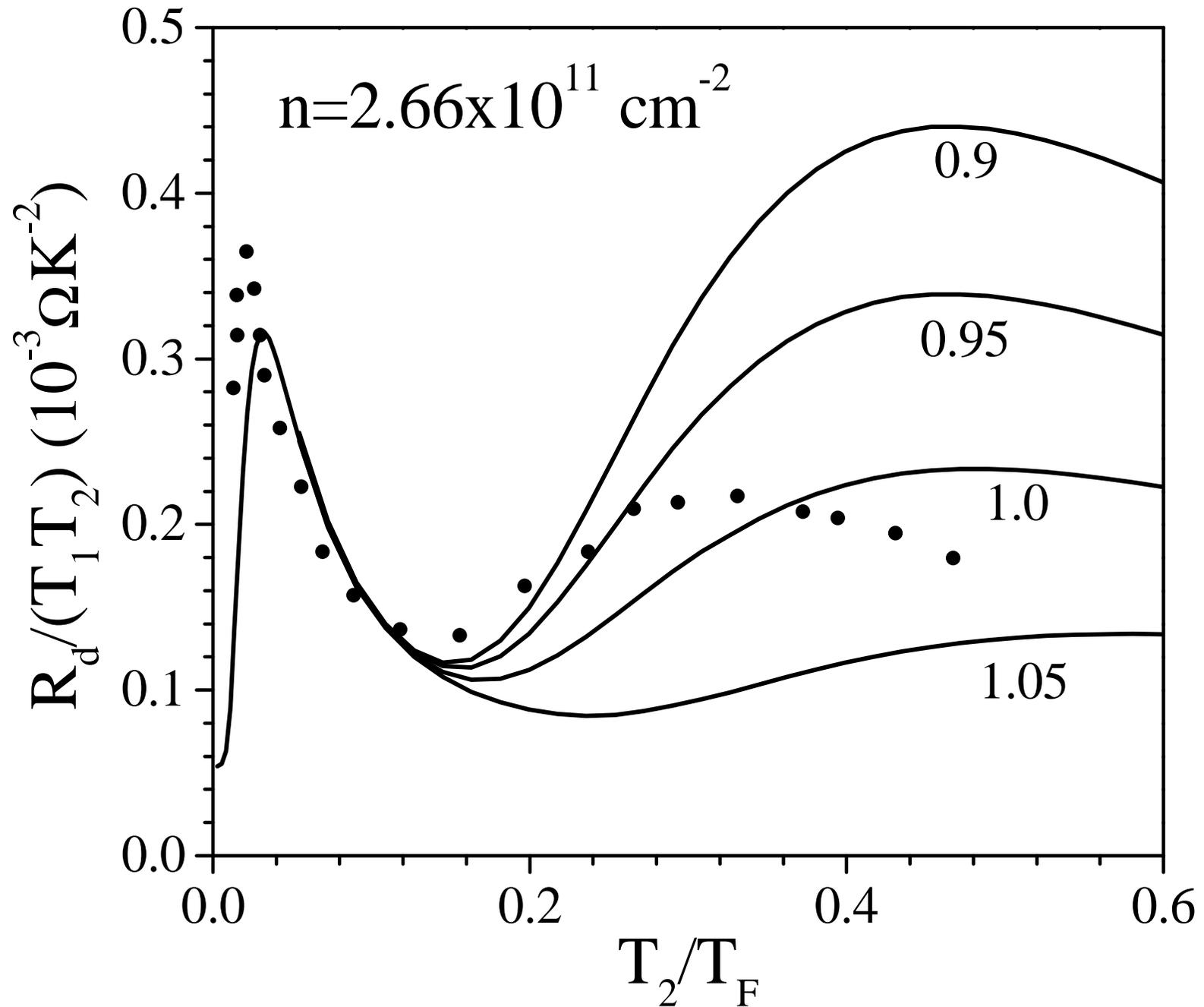
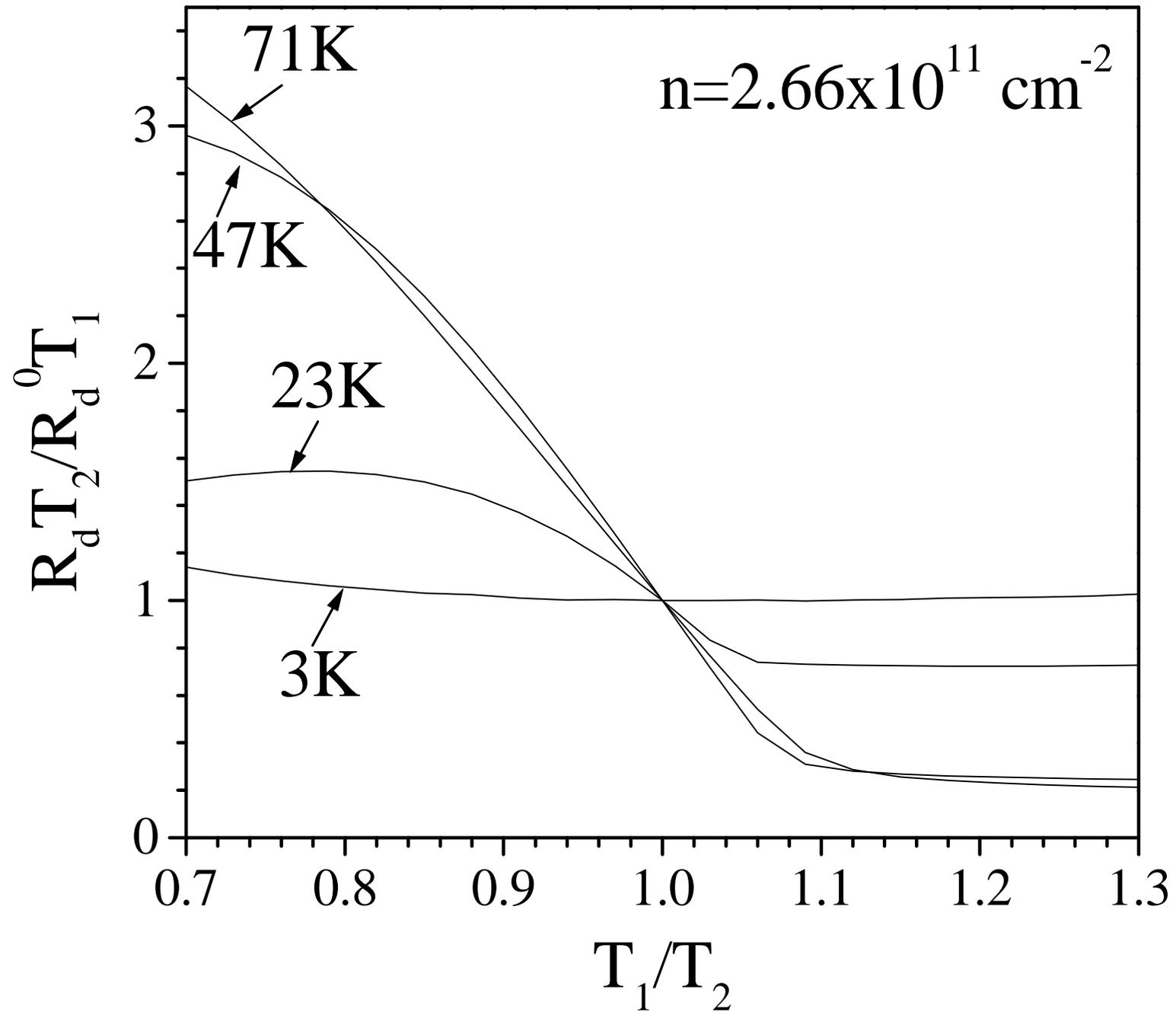
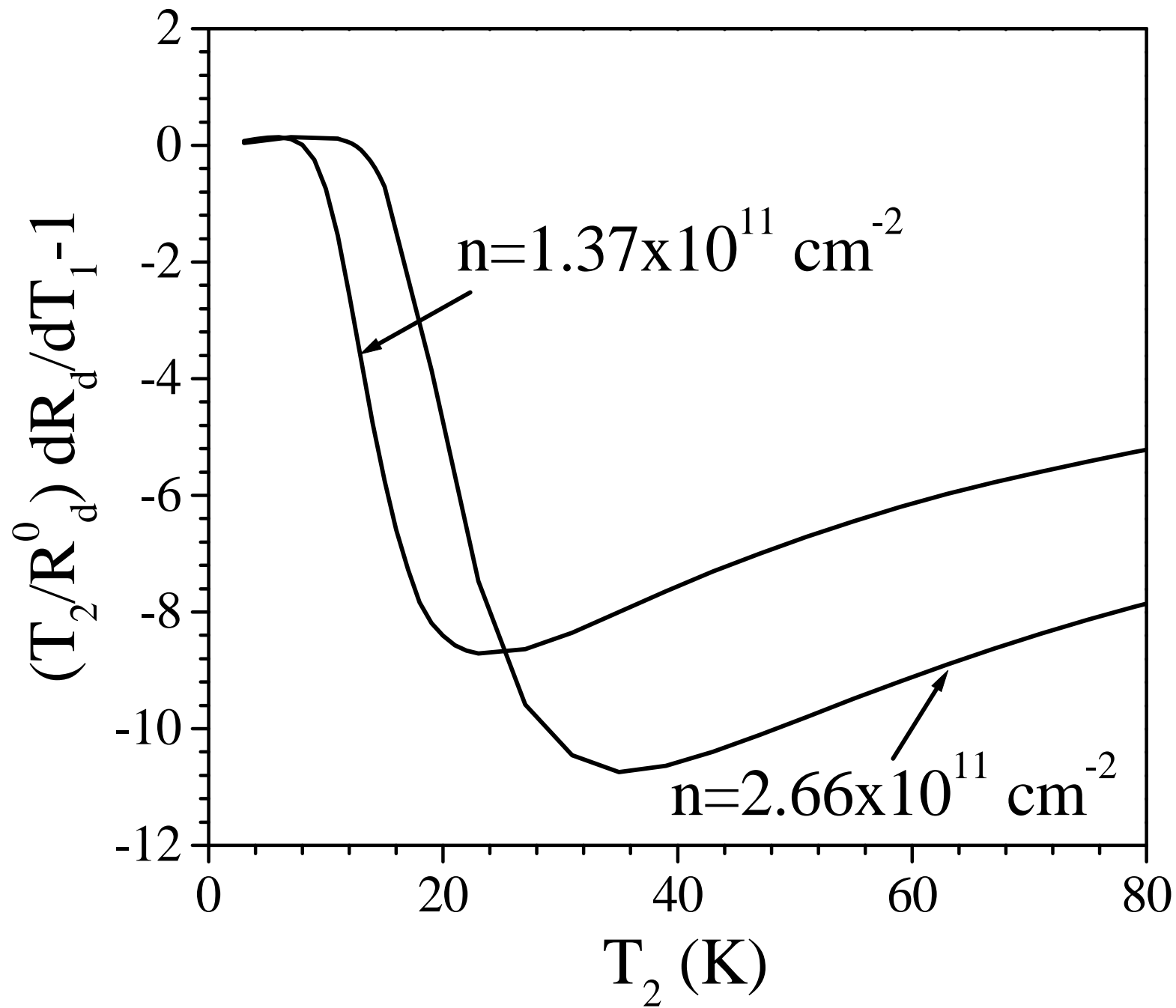
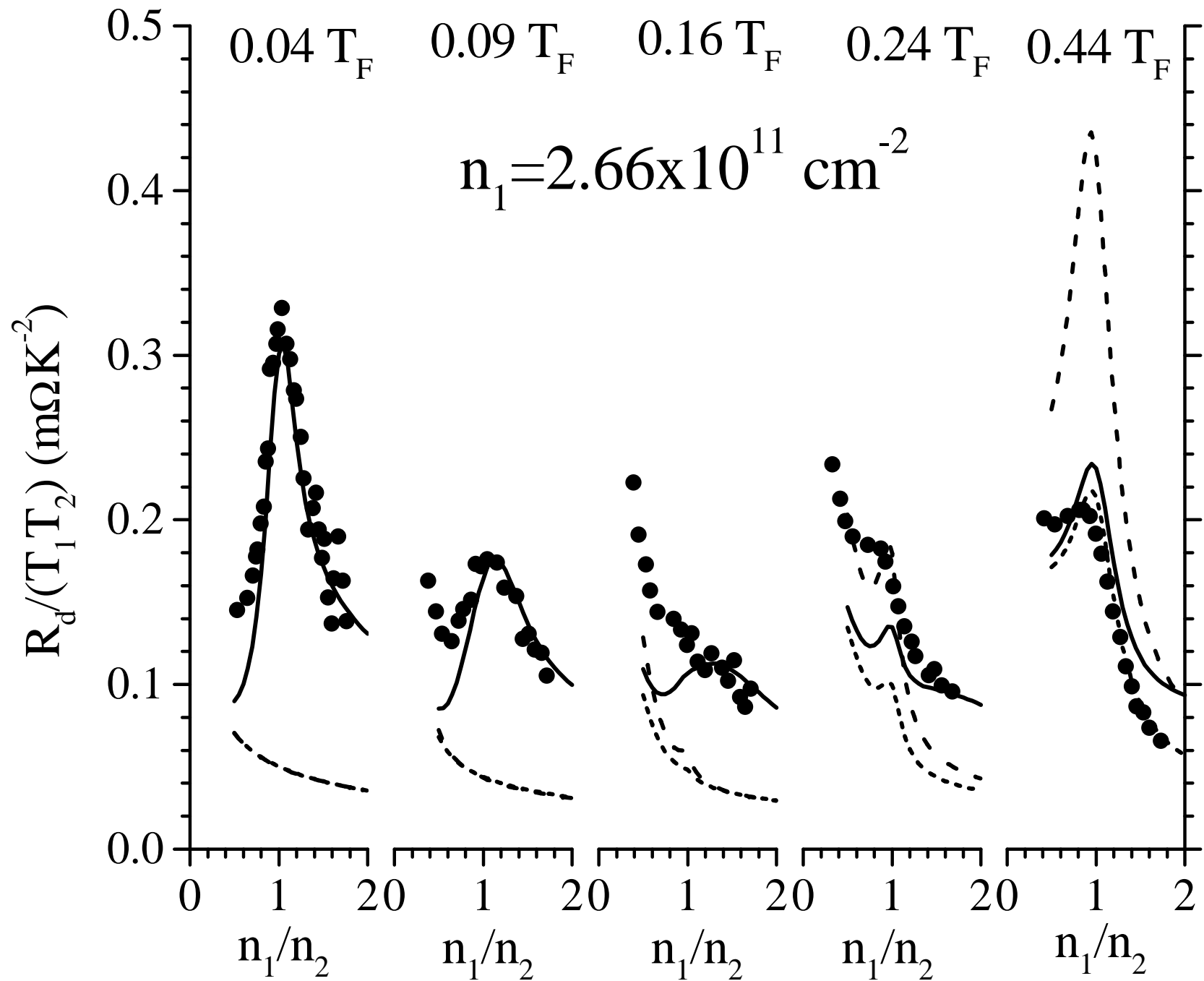
\end{document}